\documentclass[11pt]{article}

\usepackage{amssymb}
\usepackage{amsfonts}
\usepackage{amsmath}
\usepackage{amsthm}
\usepackage[english]{babel}
\usepackage{latexsym}
\usepackage{color}
\usepackage{enumerate}
\usepackage{nicefrac}
\usepackage{mathrsfs}
\usepackage{comment}
\usepackage[hmargin=2cm,vmargin=2cm]{geometry}
\usepackage{bm}
\usepackage{bbm}

\usepackage{centernot}

\usepackage[affil-it]{authblk}
\theoremstyle{plain}
\newtheorem{theorem}{Theorem}[section]
\newtheorem{lemma}[theorem]{Lemma}
\newtheorem{proposition}[theorem]{Proposition}
\newtheorem{corollary}[theorem]{Corollary}

\theoremstyle{definition}
\newtheorem{definition}[theorem]{Definition}
\newtheorem{assumption}[theorem]{Assumption}
\newtheorem{remark}[theorem]{Remark}
\newtheorem{example}[theorem]{Example}

\theoremstyle{remark}

\numberwithin{equation}{section}

\newcommand{\ba}{\begin{array}{ll}}
\newcommand{\bal}{\begin{array}{ll}}
\newcommand{\ea}{\end{array}}

\newcommand{\E}{\mathbb{E}}

\newcommand{\probp}{\mathbb{P}}
\newcommand{\probq}{\mathbb{Q}}
\newcommand{\R}{\mathbb{R}}
\newcommand{\N}{\mathbb{N}}

\newcommand{\cF}{{\mathcal{F}}}
\newcommand{\cG}{{\mathcal{G}}}

\newcommand{\cS}{{\mathcal{S}}}
\newcommand{\cA}{\mathcal{A}}
\newcommand{\cC}{\mathcal{C}}
\newcommand{\cD}{\mathcal{D}}
\newcommand{\cH}{\mathcal{H}}

\newcommand{\cP}{\mathcal{P}}

\newcommand{\cX}{{\mathcal{X}}}

\newcommand{\one}{\mathbbm 1}

\newcommand{\essinf}{\mathop {\rm ess\,inf}\nolimits}

\newcommand{\co}{\mathop{\rm co}\nolimits}

\newcommand{\cl}{\mathop{\rm cl}\nolimits}

\newcommand{\ES}{\mathop {\rm ES}\nolimits}

\def\keywords{\vspace{.5em}
{\noindent\textbf{Keywords}:\,\relax%
}}

\makeatletter
\def\@fnsymbol#1{\ensuremath{\ifcase#1\or 1\or 2\or 3\or 4\or 5\or 6\or 7\or 8\else\@ctrerr\fi}}
\makeatother

\begin{document}

\title{Law-invariant functionals on general spaces of random variables}

\author{\sc{Fabio Bellini}}
\affil{Department of Statistics and Quantitative Methods\\
University of Milano-Bicocca, Italy\\
\texttt{\normalsize fabio.bellini@unimib.it}}
\author{\sc{Pablo Koch-Medina},\,\,\sc{Cosimo Munari}}
\affil{Center for Finance and Insurance and Swiss Finance Institute\\
University of Zurich, Switzerland\\
\texttt{\normalsize pablo.koch@bf.uzh.ch},\,\,\texttt{\normalsize cosimo.munari@bf.uzh.ch}}
\author{\sc{Gregor Svindland}}
\affil{Institute of Probability and Statistics and House of Insurance\\
Leibniz University Hannover, Germany\\
\texttt{\normalsize gregor.svindland@insurance.uni-hannover.de}}

\date{\today}

\maketitle

\begin{abstract}
\noindent We establish general versions of a variety of results for quasiconvex, lower-semicontinuous, and law-invariant functionals. Our results extend well-known results from the literature to a large class of spaces of random variables. We sometimes obtain sharper versions, even for the well-studied case of bounded random variables. Our approach builds on two fundamental structural results for law-invariant functionals: the equivalence of law invariance and Schur convexity, i.e., monotonicity with respect to the convex stochastic order, and the fact that a law-invariant functional is fully determined by its behaviour on bounded random variables. We show how to apply these results to provide a unifying perspective on the literature on law-invariant functionals, with special emphasis on quantile-based representations, including Kusuoka representations, dilatation monotonicity, and infimal convolutions.
\end{abstract}


\parindent 0em \noindent


\section{Introduction}

The goal of this paper is to provide a unifying perspective on and generalizations of some of the key results on law-invariant functionals defined on spaces of random variables. Recall that a functional on a space of random variables is called law invariant if it assigns the same value to random variables that have the same probability law. Law invariance is often encountered across the finance and economics literature, e.g.\ in capital adequacy, risk sharing, portfolio selection, capital allocation, insurance pricing. Many important results on law-invariant functionals are obtained under the additional assumption of (quasi)convexity and (semi)continuity. Examples include the link between law invariance, Schur convexity, and dilatation monotonicity, see Dana \cite{Dana2005}, Cherny and Grigoriev \cite{ChernyGrigoriev2007}, Grechuk and Zabarankin \cite{GrechukZabarankin2012}, Svindland~\cite{Svindland2014}, Rahsepar and Xanthos~\cite{RahseparXanthos2020}; the quantile-based dual representation of law-invariant functionals, see Kusuoka \cite{Kusuoka2001}, Frittelli and Rosazza Gianin \cite{FrittelliRosazza2005}, Shapiro \cite{Shapiro2013}, Pichler and Shapiro \cite{PichlerShapiro2015}; the extension result for law-invariant functionals, see Filipovi\'{c} and Svindland \cite{FilipovicSvindland2012}, Chen et al.~\cite{ChenGaoXanthos2018}; the results on infimal convolutions of law-invariant functionals, see Jouini et al.~\cite{JouiniSchachermayerTouzi2008}, Filipovi\'{c} and Svindland \cite{FilipovicSvindland2008}, Ludkovski and R\"{u}schendorf \cite{LudkovskiRueschendorf2008}, Acciaio~\cite{Acciaio2009}, Ravanelli and Svindland \cite{RavanelliSvindland2014}, Chen et al.~\cite{ChenGaoXanthos2018}, Liebrich and Svindland~\cite{LiebrichSvindland2019}, Liu et al.~\cite{LiuWangWei2020}; the automatic continuity results for law-invariant functionals, see Jouini et al.\ \cite{JouiniSchachermayerTouzi2006}, Svindland \cite{Svindland2010}, Gao et al.\ \cite{GaoLeungMunariXanthos2018}, Leung and Tantrawan \cite{LeungTantrawan2020}; the link between finiteness and statistical robustness of law-invariant functionals, see Koch-Medina and Munari \cite{KochMunari2014} and Kr\"{a}tschmer et al.\ \cite{KraetschmerSchiedZaehle2014}. Most of these results have been established for law-invariant functionals defined on Lebesgue spaces with the space of bounded random variables being the most usual choice. In some cases, only minor adjustments are required to make the proof work in a more general setting. However, on other occasions, specific features of Lebesgue spaces are used making it difficult to assess whether greater generality can be achieved. This problem arises, for example, when trying to extend results from the class of Lebesgue spaces to the larger class of Orlicz spaces or to the even broader class of rearrangement-invariant spaces as discussed, e.g., in Gao et al.\ \cite{GaoLeungMunariXanthos2018}, Chen et al.~\cite{ChenGaoXanthos2018}, Liebrich and Svindland \cite{LiebrichSvindland2019}, Leung and Tantrawan~\cite{LeungTantrawan2020}. We also refer to Liebrich and Svindland \cite{LiebrichSvindland2017} for a general discussion on ``natural'' model spaces with a focus on risk measure theory. The extension of some of the above results to spaces of random vectors is discussed, e.g., in R\"{u}schendorf~\cite{Rueschendorf2006} and Ekeland and Schachermayer \cite{EkelandSchachermayer2011}.

\medskip

In this paper we deal with law-invariant functionals defined on a locally convex space $\cX$ of random variables (over a nonatomic probability space) that contains $L^\infty$ and is contained in $L^1$. For the type of results we are targeting, it is necessary that the functional in question admits a special type of dual representation. This naturally leads us to focus on ``admissible'' functionals, namely functionals that are quasiconvex and lower semicontinuous with respect to a locally convex topology on $\cX$ such that the resulting topological dual also contains $L^\infty$ and is contained in $L^1$. As we illustrate in Section~2, this setup provides sufficient flexibility to accommodate virtually all interesting cases. Our approach builds on a careful analysis of the interplay between concepts and results from probability theory --- quantile functions, stochastic orders, martingale convergence --- and convex duality. This analysis yields a series of structural results for admissible law-invariant functionals that provide a unifying perspective on and allow proving generalizations of some of the key results mentioned above in a more direct and, hopefully, insightful way.

\medskip

The first structural result is Theorem~\ref{theo: equivalence law invariance and schur convexity} establishing that, for an admissible functional, law invariance and Schur convexity, i.e., monotonicity with respect to the convex order, are equivalent. This is a generalization of an important result first proved by Dana~\cite{Dana2005} for the space of bounded random variables. The key behind this equivalence is contained in Lemma~\ref{lem: orbits}, which provides an extension of classical results by Ryff \cite{Ryff1965} and Luxemburg~\cite{Luxemburg1967} uncovering the strong topological link between law invariance and the convex order: For any random variable $X\in\cX$, the set of random variables in $\cX$ that are preferred to $X$ with respect to the convex order coincides with the closed convex hull (with respect to the chosen weak topology) of the set of random variables in $\cX$ having the same probability law as $X$.

\medskip

The link with Schur convexity is used to prove a second set of structural results starting with Theorem~\ref{theo: restriction}, which shows that an admissible law-invariant functional is fully determined by its behaviour on $L^\infty$. This ``reduction principle'' can be viewed as the main result of the paper and has a number of consequences. Most importantly, a simple application of the reduction principle to the conjugate (in the sense of quasiconvex duality) of an admissible law-invariant functional delivers a generalization of a result obtained by Svindland~\cite{Svindland2010} for $L^\infty$. Indeed, Theorem~\ref{theo: bounded dual} establishes that every admissible law-invariant functional is automatically $\sigma(\cX,L^\infty)$-lower semicontinuous. In particular, this shows that in the dual representation of an admissible law-invariant functional we can restrict our attention to dual elements belonging to $L^\infty$. This important topological property implies Proposition~\ref{prop: extension}, which states that every admissible law-invariant functional defined on $L^\infty$ can be uniquely extended to a functional of the same type on $\cX$. This provides a general formulation of the important extension result obtained by Filipovi\'{c} and Svindland in \cite{FilipovicSvindland2012}. In combination with the reduction principle, this extension result is a powerful tool to ``lift'' results on law-invariant functionals from $L^\infty$ to our general space $\cX$.

\medskip

By relying on the above general results we are able to extend a number of well-known theorems on law-invariant functionals from the literature. These include a variety of quantile-based dual representations and results on infimal convolutions and dilatation monotonicity. For some of them, exploiting the equivalence of law invariance and Schur convexity, we can provide simple direct proofs and, sometimes, sharper versions, even for the $L^\infty$ case. We also note that, in contrast to a large part of the literature, we consistently work with quasiconvex, instead of only convex, functionals. We refer to the individual sections for a detailed comparison with the literature.

\medskip

The paper is organized as follows. After describing the setting in Section 2, we study the relation between law invariance and Schur convexity in Section 3. Based on this, we prove the fundamental reduction principle for law-invariant functionals in Section 4. All the aforementioned applications are discussed in Section 5.


\section{Setting}
\label{sect: setup}

Let $(\Omega,\cF,\probp)$ be a nonatomic probability space. Any Borel measurable function $X:\Omega\to\R$ is called a random variable. We denote by $L^0$ the set of equivalent classes of random variables with respect to almost-sure equality (under $\probp$). As usual, we never explicitly distinguish between an element of $L^0$ and any of its representatives. Equalities and inequalities between random variables are always meant in the almost-sure sense. In particular, the elements of $\R$ are identified with random variables that are almost-surely constant. For every set $\cX\subset L^0$ we denote by $\cX_+$ the set of positive random variables in $\cX$. The standard Lebesgue spaces on $(\Omega,\cF,\probp)$ are denoted by $L^p$ for $p\in[1,\infty]$. The expectation under $\probp$ is simply denoted by $\E$.

\begin{definition}
For two random variables $X,Y\in L^0$ we write $X\sim Y$ whenever $X$ and $Y$ have the same probability law under $\probp$. A set $\cX\subset L^0$ is said to be {\em law invariant} if for all $X,Y\in L^0$ we have
\[
X\in\cX, \ Y\sim X \ \implies \ Y\in\cX.
\]
A functional $\varphi:\cX\to[-\infty,\infty]$ is called {\em law invariant} if for all $X,Y\in\cX$ we have
\[
X\sim Y \ \implies \ \varphi(X)=\varphi(Y).
\]
\end{definition}

\smallskip

Throughout the entire paper we work with the following pair of linear subspaces of $L^0$.

\begin{assumption}
\label{assumption one}
Let $\cX$ and $\cX^\ast$ be two linear subspaces of $L^0$ such that:
\begin{enumerate}[(1)]
  \item $\cX$ and $\cX^\ast$ are law-invariant.
  \item $XY\in L^1$ for all $X\in\cX$ and $Y\in\cX^\ast$.
  \item $L^\infty\subseteq\cX\subseteq L^1$ and $L^\infty\subseteq\cX^\ast\subseteq L^1$.
\end{enumerate}
We denote by $\sigma(\cX,\cX^\ast)$ the weakest topology on $\cX$ with respect to which the linear functional $\varphi_Y:\cX\to\R$ given by $\varphi_Y(X):=\E[XY]$ is continuous for every $Y\in\cX^\ast$. Equipped with this topology, $\cX$ becomes a locally convex Hausdorff topological vector space (see e.g.\ \cite{Zalinescu2002}).
\end{assumption}

\medskip

The following example shows concrete instances of pairs satisfying the above assumption.

\begin{example}
\label{ex Orlicz}
Let $\Phi:[0,\infty)\to[0,\infty]$ be an Orlicz function, i.e.\ a convex, left-continuous, increasing function which is finite on a right neighborhood of zero and satisfies $\Phi(0)=0$. The conjugate of $\Phi$ is the function $\Phi^\ast:[0,\infty)\to[0,\infty]$ defined by
\[
\Phi^\ast(u) := \sup_{t\in[0,\infty)}\{tu-\Phi(t)\}.
\]
Note that $\Phi^\ast$ is also an Orlicz function. For every $X\in L^0$ define the Luxemburg norm by
\[
\|X\|_\Phi := \inf\left\{\lambda\in(0,\infty) \,; \ \E\left[\Phi\left(\frac{|X|}{\lambda}\right)\right]\leq1\right\}.
\]
The corresponding Orlicz space is given by
\[
L^\Phi:=\{X\in L^0 \,; \ \|X\|_\Phi<\infty\}.
\]
The heart of $L^\Phi$ is the space
\[
H^ \Phi:=\left\{X\in L^\Phi \,; \ \forall \lambda\in(0,\infty) \,:\, \E\left[\Phi\left(\frac{|X|}{\lambda}\right)\right]<\infty\right\}.
\]
The classical Lebesgue spaces are prominent examples of Orlicz spaces. On the one side, setting $\Phi(t)=t^p$ for $p\in[1,\infty)$ and $t\in[0,\infty)$, we have $L^\Phi=H^\Phi=L^p$ and the Luxemburg norm coincides with the usual $p$ norm. On the other side, if we define $\Phi(t)=0$ for $t\in[0,1]$ and $\Phi(t)=\infty$ otherwise, then we have $L^\Phi=L^\infty$ and the Luxemburg norm coincides with the usual esssup norm. Note that, in this case, $H^\Phi=\{0\}$.

\smallskip

In our nonatomic setting we have $L^\Phi=H^\Phi$ if and only if $\Phi$ is $\Delta_2$, i.e.\ there exist $s\in(0,\infty)$ and $k\in(0,\infty)$ such that $\Phi(2t)<k\Phi(t)$ for every $t\in[s,\infty)$. A well-known example of a nontrivial $H^\Phi$ with $H^\Phi\neq L^\phi$ corresponds to the choice $\Phi(t)=\exp(t)-1$ for $t\in[0,\infty)$.

\smallskip

The norm dual of $L^\Phi$ cannot be identified with a subspace of $L^0$ in general. However, if $\Phi$ is finite valued (otherwise $H^\Phi=\{0\}$), the norm dual of $H^\Phi$ can always be identified with $L^{\Phi^\ast}$. For the case $L^p$, for $p\in[1,\infty)$, this is simply the well-known identification of the norm dual of $L^p$ with $L^{\frac{p}{p-1}}$ (with the usual convention $\frac10:=\infty$). For more details on Orlicz spaces we refer to \cite{EdgarSucheston1992}.

\smallskip

The following pairs satisfy Assumption~\ref{assumption one} (we assume that none of the Orlicz hearts is reduced to zero):
\begin{enumerate}[(1)]
  \item $\cX=L^\Phi$ and $\cX^\ast\in\{L^{\Phi^\ast},H^{\Phi^\ast},L^\infty\}$.
  \item $\cX=H^\Phi$ and $\cX^\ast\in\{L^{\Phi^\ast},H^{\Phi^\ast},L^\infty\}$.
\end{enumerate}
\end{example}

\smallskip

We conclude this section by recalling some fundamental notions about functionals. We work under the above standing assumption. The focus of the paper is on functionals of the form
\[
\varphi:\cX\to[-\infty,\infty].
\]
We say that $\varphi$ is {\em proper} if $\varphi(X)>-\infty$ for every $X\in\cX$ and $\varphi(X)<\infty$ for some $X\in\cX$. We say that $\varphi$ is {\em increasing}, respectively {\em decreasing} if it is increasing, respectively decreasing, with respect to the almost-sure partial order, i.e.\ for all $X,Y\in\cX$ we have
\[
X\leq Y \ \implies \ \varphi(X)\leq\varphi(Y), \ \ \ \mbox{resp.\ $\varphi(X)\geq\varphi(Y)$}.
\]
The functional $\varphi$ is said to be {\em convex} if for all $X,Y\in\cX$ and $\lambda\in[0,1]$ we have
\[
\varphi(\lambda X+(1-\lambda)Y)\leq\lambda\varphi(X)+(1-\lambda)\varphi(Y)
\]
and {\em quasiconvex} if for all $X,Y\in\cX$ and $\lambda\in[0,1]$ we have
\[
\varphi(\lambda X+(1-\lambda)Y)\leq\max\{\varphi(X),\varphi(Y)\}.
\]
Clearly, every convex functional is automatically quasiconvex. Finally, $\varphi$ is called {\em $\sigma(\cX,\cX^\ast)$-lower semicontinuous} if for every net $(X_\nu)\subset\cX$ and every $X\in\cX$ we have
\[
X_\nu\xrightarrow{\sigma(\cX,\cX^\ast)}X \ \implies \ \varphi(X)\leq\liminf_\nu\varphi(X_\nu).
\]
As is well-known, $\varphi$ is quasiconvex, respectively $\sigma(\cX,\cX^\ast)$-lower semicontinuous, if and only if the set
\[
\{\varphi\leq\alpha\} := \{X\in\cX \,; \ \varphi(X)\leq\alpha\}
\]
is convex, respectively $\sigma(\cX,\cX^\ast)$-closed, for every $\alpha\in\R$. The property of $\sigma(\cX,\cX^\ast)$-lower semicontinuity will play a key role in the sequel. The next proposition features three operative characterizations of this property in the common situation where the ambient space $\cX$ can be equipped with a rearrangement-invariant structure. Recall that $\cX$ is a {\em rearrangement-invariant space} if it is a solid lattice (with respect to the almost-sure partial order) equipped with a law-invariant, complete, lattice norm; see \cite{Luxemburg1967}. Note that any Luxemburg norm has the Lebesgue property if it is restricted to an Orlicz heart.

\begin{definition}
We say that a functional $\varphi:\cX\to[-\infty,\infty]$ has the {\em Fatou property} if for every sequence $(X_n)\subset\cX$ and every $X\in\cX$ we have
\[
X_n\xrightarrow{a.s.}X, \ \sup_{n\in\N}|X_n|\in\cX \ \implies \ \varphi(X)\leq\liminf_{n\to\infty}\varphi(X_n).
\]
Similarly, $\varphi$ has the {\em Lebesgue property} if for every sequence $(X_n)\subset\cX$ and every $X\in\cX$ we have
\[
X_n\xrightarrow{a.s.}X, \ \sup_{n\in\N}|X_n|\in\cX \ \implies \ \varphi(X)=\lim_{n\to\infty}\varphi(X_n).
\]
We say that $\varphi$ is {\em continuous from below} if for every sequence $(X_n)\subset\cX$ and every $X\in\cX$ we have
\[
X_n\downarrow X \ \implies \ \varphi(X)=\lim_{n\to\infty}\varphi(X_n).
\]
\end{definition}

\smallskip

\begin{proposition}
\label{prop: fatou}
Assume that $\cX$ is a rearrangement-invariant space. Let $\varphi:\cX\to[-\infty,\infty]$ be quasiconvex and law invariant. Then, the following statements are equivalent:
\begin{enumerate}
  \item[(a)] $\varphi$ is $\sigma(\cX,\cX^\ast)$-lower semicontinuous.
  \item[(b)] $\varphi$ has the Fatou property.
\end{enumerate}
If either $\cX=L^\infty$ or $\|\cdot\|$ has the Lebesgue property, then (a) is also equivalent to:
\begin{enumerate}
  \item[(c)] $\varphi$ is norm-lower semicontinuous.
\end{enumerate}
If $\varphi$ is increasing, then (a) is also equivalent to:
\begin{enumerate}
  \item[(d)] $\varphi$ is continuous from below.
\end{enumerate}
\end{proposition}
\begin{proof}
Note that any functional with the Fatou property is automatically norm-lower semicontinuous. This is because every norm-convergent sequence admits a dominated subsequence that converges almost surely. Note also that any norm-lower semicontinuous functional satisfies the Fatou property provided that the norm has the Lebesgue property. It is also well known that the Fatou property and continuity from below are equivalent for an increasing functional. The remaining implications follow from \cite[Proposition 1.1]{Svindland2010} (see also \cite[Theorem 3.2]{Delbaen2002} and \cite[Theorem 2.1]{JouiniSchachermayerTouzi2006}) in the bounded setting, from \cite[Theorem 1.1]{GaoLeungMunariXanthos2018} in the Orlicz setting, and from \cite[Proposition 2.11]{ChenGaoXanthos2018} and \cite[Theorem 3.1]{LeungTantrawan2020} in the general setting. We also refer to \cite{BiaginiFrittelli2009} for early results linking lower semicontinuity with the Fatou property.
\end{proof}

\smallskip

We conclude this section by recalling the classical Fenchel-Moreau dual representations in our setting. To this effect, recall that the {\em conjugate} of $\varphi$ is the functional $\varphi^\ast:\cX^\ast\to[-\infty,\infty]$ defined by
\[
\varphi^\ast(Y) := \sup_{X\in\cX}\{\E[XY]-\varphi(X)\}.
\]
Moreover, we define $F_\varphi^\ast:\cX^\ast\times\R\to[-\infty,\infty]$ by
\[
F_\varphi^\ast(Y,\alpha) := \inf\{\varphi(X) \,; \ X\in\cX, \ \E[XY]\geq\alpha\}.
\]
The above functionals appear in the classical Fenchel-Moreau dual representations, see e.g.\ \cite[Theorem 2.3.3]{Zalinescu2002} and \cite[Theorem 2.6, Lemma 2.7]{FrenkDiasGromicho1994}.

\begin{proposition}
\label{lem: fenchel moreau}
Let $\varphi:\cX\to[-\infty,\infty]$ be proper and $\sigma(\cX,\cX^\ast)$-lower semicontinuous.
\begin{enumerate}[(i)]
  \item If $\varphi$ is convex, then
\[
\varphi(X) = \sup_{Y\in\cX^\ast}\{\E[XY]-\varphi^\ast(Y)\}, \ \ \ X\in\cX.
\]
  \item If $\varphi$ is quasiconvex, then
\[
\varphi(X) = \sup_{Y\in\cX^\ast}F_\varphi^\ast(Y,\E[XY]), \ \ \ X\in\cX.
\]
\end{enumerate}
If $\varphi$ is additionally increasing, then we can replace $\cX^\ast$ by $\cX^\ast_+$ in the above statements.
\end{proposition}


\section{Law invariance and Schur convexity}

In this section we start our study of law-invariant functionals on general spaces of random variables by showing the link between law invariance and stochastic orders. More precisely, we establish that, for a quasiconvex and lower semicontinuous functional, law invariance is equivalent to monotonicity with respect to the convex order. Throughout the whole section we maintain our standing assumption from Section~\ref{sect: setup}.

\medskip

As a preliminary step, we recall the notion of a quantile function and establish a useful Hardy-Littlewood type result for the pair $(\cX,\cX^\ast)$, which highlights the link between quantile functions and law invariance. This result plays an important role in the literature, where it is typically used in the context of the pairs $(L^1,L^\infty)$ and $(L^\infty,L^1)$, see, e.g., \cite{Dana2005}. We show that everything goes through in our general setting because we can easily ensure the integrability of the underlying quantile functions and of their products.

\begin{definition}
For every random variable $X\in L^0$ we denote by $q_X$ a fixed but arbitrary quantile function for $X$, i.e.\ a function $q_X:(0,1)\to\R$ satisfying
\[
\inf\{x\in\R \,; \ \probp(X\leq x)\geq s\} \leq q_X(s) \leq \inf\{x\in\R \,; \ \probp(X\leq x)>s\}
\]
for every $s\in(0,1)$. Note that, since the cumulative distribution function of $X$ has at most countably many discontinuity points, any two quantile functions for $X$ coincide almost surely with respect to the Lebesgue measure on $(0,1)$.
\end{definition}

\smallskip

\begin{lemma}
\label{lem: chong rice}
For all $X\in\cX$ and $Y\in\cX^\ast$ the functions $q_{X}(1-\cdot)q_{Y}$ and $q_{X}q_{Y}$ are Lebesgue integrable on $(0,1)$ and the following statements hold:
\begin{enumerate}[(i)]
  \item $\inf_{X'\sim X}\E[X'Y] = \inf_{Y'\sim Y}\E[XY'] = \int_0^1q_X(1-s)q_Y(s)ds$.
  \item $\sup_{X'\sim X}\E[X'Y] = \sup_{Y'\sim Y}\E[XY'] = \int_0^1q_X(s)q_Y(s)ds$.
\end{enumerate}
In addition, all infima and suprema above are attained.
\end{lemma}
\begin{proof}
Take $X\in\cX$ and $Y\in\cX^\ast$. Since $(\Omega,\cF,\probp)$ is nonatomic, it follows from~\cite[Theorem 2.6]{BennettSharpley1988} that
\[
0 \leq \int_0^1q_{|X|}(s)q_{|Y|}(s)ds = \E[|XY'|]
\]
for some $Y'\sim Y$. The law invariance of $\cX^\ast$ implies that $Y'\in\cX^\ast$ so that $XY'\in L^1$. This shows that $q_{|X|}q_{|Y|}$ is Lebesgue integrable on $(0,1)$. Now, it is not difficult to verify, see e.g.\ \cite[Theorem 4.6]{ChongRice1971}, that
\[
|q_Xq_Y| = q_{\max\{X,0\}}q_{\max\{Y,0\}}+q_{\min\{X,0\}}q_{\min\{Y,0\}}-q_{\max\{X,0\}}q_{\min\{Y,0\}}-q_{\min\{X,0\}}q_{\max\{Y,0\}} \leq 4q_{|X|}q_{|Y|}.
\]
This yields the Lebesgue integrability of $q_{X}q_{Y}$. As $q_X(1-\cdot)=-q_{-X}$ almost surely with respect to the Lebesgue measure on $(0,1)$, this also yields the Lebesgue integrability of $q_{X}(1-\cdot)q_{Y}$. The remaining statements follow from \cite[Theorem 13.4]{ChongRice1971}.
\end{proof}

\smallskip

After recalling the notion of convex order, we provide a number of equivalent conditions, expressed in terms of quantile functions, for two random variables to be ranked in the convex order. In particular, assertion {\em (c)} states the well-known link between the convex order and second-order stochastic dominance. This provides a minor extension of the results in \cite{Dana2005}.

\begin{definition}
For all random variables $X,Y\in L^1$ we write $X\succeq_{cx}Y$ whenever $X$ dominates $Y$ in the {\em convex order}, i.e. whenever
\[
\mbox{$\E[f(X)]\geq\E[f(Y)]$ for every convex function $f:\R\to\R$}.
\]
A set $\cA\subset\cX$ is {\em Schur convex} if for all $X,Y\in\cX$ we have
\[
X\in\cA, \ X\succeq_{cx}Y \ \implies \ Y\in\cA.
\]
Similarly, a functional $\varphi:\cX\to[-\infty,\infty]$ is {\em Schur convex} if for all $X,Y\in\cX$ we have
\[
X\succeq_{cx}Y \ \implies \ \varphi(X)\geq\varphi(Y).
\]
\end{definition}

\smallskip

\begin{lemma}
\label{lem: characterization convex order}
For all $X,Y\in L^1$ the following statements are equivalent:
\begin{enumerate}[(a)]
  \item For every nondecreasing function $g:(0,1)\to\R$ such that $q_Xg$ and $q_Yg$ are Lebesgue integrable on $(0,1)$ we have
\[
\int_0^1 q_X(s)g(s)ds \geq \int_0^1 q_Y(s)g(s)ds.
\]
  \item For every nondecreasing bounded function $g:(0,1)\to\R$ we have
\[
\int_0^1 q_X(s)g(s)ds \geq \int_0^1 q_Y(s)g(s)ds.
\]
  \item $\E[X]=\E[Y]$ and for every $s\in(0,1)$
\[
\int_0^s q_X(t)dt \ge \int_0^s q_Y(t)dt.
\]
  \item $X\succeq_{cx}Y$.
\end{enumerate}
\end{lemma}
\begin{proof}
It is well known that {\em (c)} and {\em (d)} are equivalent, see e.g.\ \cite[Lemma 2.1]{Dana2005}. Moreover, it follows from \cite[Lemma 2.2]{Dana2005} that {\em (b)} and {\em (d)} are equivalent. Hence, we only need to show that {\em (a)} and {\em (b)} are equivalent. It is clear that {\em (a)} implies {\em (b)} because both $q_X$ and $q_Y$ are Lebesgue integrable on $(0,1)$, see e.g.\ \cite[Proposition 4.3]{ChongRice1971}. To conclude the proof, assume that {\em (b)} holds and take an arbitrary nondecreasing function $g:(0,1)\to\R$ such that $q_Xg$ and $q_Yg$ are Lebesgue integrable on $(0,1)$. For every $n\in\N$ define a function $g_n:(0,1)\to\R$ by setting
\[
g_n(s)=
\begin{cases}
-n & \mbox{if} \ g(s)<-n,\\
g(s) & \mbox{if} \ -n\leq g(s)\leq n,\\
n & \mbox{if} \ g(s)>n.
\end{cases}
\]
By assumption, we have
\[
\int_0^1 q_X(s)g_n(s)ds \geq \int_0^1 q_Y(s)g_n(s)ds
\]
for every $n\in\N$. Since $g_n\to g$ pointwise and $|q_Xg_n|\leq|q_Xg|$ as well as $|q_Yg_n|\leq|q_Yg|$ for every $n\in\N$, it follows from the Dominated Convergence Theorem that
\[
\int_0^1 q_X(s)g(s)ds = \lim_{n\to\infty}\int_0^1 q_X(s)g_n(s)ds \geq \lim_{n\to\infty}\int_0^1 q_Y(s)g_n(s)ds = \int_0^1 q_Y(s)g(s)ds.
\]
This concludes the proof of the equivalence.
\end{proof}

\smallskip

The last ingredient to prove the announced equivalence between law invariance and Schur convexity is contained in the following density result. This shows that, for every random variable $X$ belonging to $\cX$, the set of all random variables in $\cX$ that are dominated in the convex order by $X$ coincides with the closed convex hull of the law-invariance equivalence class $\{Y\in\cX \,; \ Y\sim X\}$. The closure is taken with respect to the weak topology $\sigma(\cX,\cX^\ast)$. In particular, every random variable that is dominated by $X$ in the convex order can be approximated in said topology by a net of convex combinations of random variables that are identically distributed as $X$. This density result was first established in $L^1$ in \cite[Theorem 5]{Ryff1965} and adapted to $L^\infty$ (paired with $L^1$) in \cite[Proposition 4.1]{Dana2005} and to general $L^p$ spaces in \cite[Corollary 2.5]{Svindland2014}. Here, for a set $\cA\subset\cX$ we denote by $\co(\cA)$ its convex hull and by $\cl_{\sigma(\cX,\cX^\ast)}(\cA)$ its $\sigma(\cX,\cX^\ast)$-closure.

\begin{lemma}
\label{lem: orbits}
For every $X\in\cX$ we have $\{Y\in\cX \,; \ X\succeq_{cx}Y\} = \cl_{\sigma(\cX,\cX^\ast)}(\co(\{Y\in\cX \,; \ Y\sim X\}))$.
\end{lemma}
\begin{proof}
For convenience we use the following notation throughout the proof:
\[
\cC(X)=\{Y\in L^1 \,; \ X\succeq_{cx}Y\}, \ \ \ \cD(X)=\{Y\in L^1 \,; \ Y\sim X\}=\{Y\in\cX \,; \ Y\sim X\}.
\]
To establish the inclusion ``$\subset$'', assume there exists $Y\in\cC(X)\cap\cX$ outside of $\cl_{\sigma(\cX,\cX^\ast)}(\co(\cD(X)))$. Then, by Hahn-Banach Separation, we find $Z\in\cX^\ast$ such that $\sup\{\E[X'Z] \,; \ X'\sim X\}<\E[YZ]$. It now follows from Lemma~\ref{lem: chong rice} and Lemma~\ref{lem: characterization convex order} that
\[
\int_0^1q_X(s)q_Z(s)ds < \E[YZ] \leq \int_0^1q_Y(s)q_Z(s)ds \leq \int_0^1q_X(s)q_Z(s)ds,
\]
where we used that $X\succeq_{cx}Y$. This delivers a contradiction and yields the desired inclusion.

\smallskip

To establish the inclusion ``$\supset$'', note first that $\cC(X)\cap\cX$ is convex, law invariant, and contains $\cD(X)$. Hence, it suffices to prove that $\cC(X)\cap\cX$ is $\sigma(\cX,\cX^\ast)$-closed. To this effect, note that $\cX$, equipped with the topology $\sigma(\cX,\cX^\ast)$, is continuously embedded into $L^1$, equipped with the topology $\sigma(L^1,L^\infty)$. This is because $L^\infty$ is contained in $\cX^\ast$ by our standing assumption. As $\cC(X)$ is $\sigma(L^1,L^\infty)$-compact by \cite[Corollary~3.2]{Chong1974}, we infer that $\cC(X)\cap\cX$ is $\sigma(\cX,\cX^\ast)$-closed.
\end{proof}

\smallskip

We come to the announced equivalence between law invariance and Schur convexity. In the convex case, this result extends \cite[Theorem 4.1]{Dana2005}, which was established in $L^\infty$, and \cite[Proposition 2.4]{GrechukZabarankin2012}, which was established in a general $L^p$ space under the additional assumption of positive homogeneity. In the quasiconvex case, the result extends \cite[Theorem 5.1]{CerreiaMaccheroniMarinacciMontrucchio2011}, which was established in $L^\infty$ under the additional assumption of monotonicity, and \cite[Theorem 18]{LiebrichSvindland2019}, which was established in the setting of a rearrangement-invariant space.

\begin{theorem}
\label{theo: equivalence law invariance and schur convexity}
For a proper, quasiconvex, $\sigma(\cX,\cX^\ast)$-lower semicontinuous functional $\varphi:\cX\to[-\infty,\infty]$ the following statements are equivalent:
\begin{enumerate}[(a)]
  \item $\varphi$ is law invariant.
  \item $\varphi$ is Schur convex.
\end{enumerate}
\end{theorem}
\begin{proof}
It is clear that Schur convexity implies law invariance. Now, assume that $\varphi$ is law invariant and take two arbitrary $X,Y\in\cX$ such that $X\succeq_{cx}Y$. By Lemma~\ref{lem: orbits}, the random variable $Y$ is the $\sigma(\cX,\cX^\ast)$-limit of a net $(Y_\nu)\subset\cX$ such that
\[
Y_\nu = \sum_{i=1}^{n_\nu}\lambda^\nu_i X^\nu_i
\]
for $n_\nu\in\N$ and for suitable $X^\nu_1,\dots,X^\nu_{n_\nu}\sim X$ and $\lambda^\nu_1,\dots,\lambda^\nu_{n_\nu}\in[0,1]$ summing up to one. Note that by law invariance and quasiconvexity we have
\[
\varphi(Y_\nu) \leq \max\{\varphi(X_1^\nu),\dots,\varphi(X_{n_\nu}^\nu)\} = \varphi(X)
\]
for every $\nu$. Hence, $\sigma(\cX^\ast,\cX)$-lower semicontinuity yields
\[
\varphi(X) \geq \liminf_\nu\varphi(Y_\nu) \geq \varphi(Y).
\]
This establishes that $\varphi$ is Schur convex.
\end{proof}

\smallskip

A companion result to Theorem~\ref{theo: equivalence law invariance and schur convexity} for sets establishes that law invariance is equivalent to monotonicity with respect to the convex order. The statement follows immediately by applying the above result to indicator functionals.

\begin{corollary}
\label{cor: equivalence law invariance schur sets}
For a convex and $\sigma(\cX,\cX^\ast)$-closed set $\cA\subset\cX$ the following statements are equivalent:
\begin{enumerate}[(a)]
  \item $\cA$ is law invariant.
  \item $\cA$ is Schur convex.
\end{enumerate}
\end{corollary}

\section{Restriction results}

In this section we unveil the deep link between law invariance and boundedness. We shed light on this link from a primal and a dual perspective. On the one side, we show that a quasiconvex and lower semicontinuous functional that is law invariant is uniquely determined by its action on bounded random variables. On the other side, by applying this restriction result to conjugate functionals, we show that the space of bounded random variables is the natural dual space under law invariance. The key to these results is to combine the equivalence between law invariance and Schur convexity established in the preceding section with the following approximation result. Here, for every $X\in\cX$ we denote by $\sigma(X)$ the smallest $\sigma$-field in $\cF$ with respect to which $X$ is measurable. Similarly, for every $\cG\subset\cF$ we denote by $\sigma(\cG)$ the smallest $\sigma$-field in $\cF$ containing $\cG$.

\begin{lemma}
\label{lem: density bounded}
For every $X\in\cX$ and for every increasing sequence $(\cF_n(X))$ of finitely-generated $\sigma$-fields in $\cF$ such that $\sigma(X)=\sigma(\bigcup_{n\in\N}\cF_n(X))$ we have
\[
\E[X\,|\,\cF_n(X)]\xrightarrow{\sigma(\cX,\cX^\ast)}X.
\]
\end{lemma}
\begin{proof}
Let $X\in\cX$ and take an increasing sequence $(\cF_n(X))$ of finitely-generated $\sigma$-fields in $\cF$ such that $\sigma(X)=\sigma(\bigcup_{n\in\N}\cF_n(X))$. By the Martingale Convergence Theorem we have that $\E[X\,|\,\cF_n(X)]\to X$ almost surely. Now, take any $Y\in\cX^\ast$ and $E\in\cF$ and note that $q_{\one_E|Y|}\leq \one_{(1-\probp(E),1)}q_{|Y|}$ almost surely with respect to the Lebesgue measure on $(0,1)$. Since $|X|\succeq_{cx}\E[|X|\,|\,\cF_n(X)]$ for every $n\in\N$ by Jensen's Inequality, it follows from Lemma~\ref{lem: chong rice} and Lemma~\ref{lem: characterization convex order} that
\[
\E[\one_E|\E[X\,|\,\cF_n(X)]Y|] \leq
\int_0^1q_{\E[|X|\,|\,\cF_n(X)]}(s)q_{\one_E|Y|}(s)ds \leq
\int_{1-\probp(E)}^1q_{|X|}(s)q_{|Y|}(s)ds.
\]
This shows that the sequence $(\E[X\,|\,\cF_n(X)]Y)$ is uniformly integrable. As $\E[X\,|\,\cF_n(X)]Y\to XY$ almost surely, we obtain that $\E[\E[X\,|\,\cF_n(X)]Y]\to\E[XY]$, concluding the proof.
\end{proof}

\smallskip

It follows from the preceding lemma that $L^\infty$ is dense in $\cX$ with respect to the topology $\sigma(\cX,\cX^\ast)$. This should not be surprising and could be proved in a more direct way. Indeed, for every $X\in\cX$, a direct application of the Dominated Convergence Theorem shows that
\[
\one_{\{|X|\leq n\}}X\xrightarrow{\sigma(\cX,\cX^\ast)}X.
\]
What is powerful about the preceding lemma is that we can approximate every element in $\cX$ via a sequence of special bounded random variables, namely {\em conditional expectations with respect to finitely-generated $\sigma$-fields}. This allows us to exploit Schur convexity in order to establish the announced restriction result. A similar result has been obtained by way of a different argument in \cite[Lemma 5.4]{GaoLeungMunariXanthos2018} in the Orlicz setting and \cite[Corollary 2.5]{ChenGaoXanthos2018} in the setting of a rearrangement-invariant space. Here, for a functional $\varphi:\cX\to(-\infty,\infty]$ and a subset $\cA\subset\cX$ we denote by $\varphi_{|\cA}$ the restriction of $\varphi$ to $\cA$.

\begin{theorem}
\label{theo: restriction}
Let $\varphi:\cX\to[-\infty,\infty]$ be proper, quasiconvex, $\sigma(\cX,\cX^\ast)$-lower semicontinuous, and law invariant. For every $X\in\cX$ and for every increasing sequence $(\cF_n(X))$ of finitely-generated $\sigma$-fields in $\cF$ such that $\sigma(X)=\sigma(\bigcup_{n\in\N}\cF_n(X))$ we have
\[
\varphi(X) = \lim_{n\to\infty}\varphi_{|L^\infty}(\E[X\,|\,\cF_n(X)]).
\]
In particular, $\varphi$ is uniquely determined by its restriction to $L^\infty$.
\end{theorem}
\begin{proof}
Let $X\in\cX$ and take an increasing sequence $(\cF_n(X))$ of finitely-generated $\sigma$-fields in $\cF$ such that $\sigma(X)=\sigma(\bigcup_{n\in\N}\cF_n(X))$. By Lemma~\ref{lem: density bounded} we have $\E[X\,|\,\cF_n(X)]\to X$ with respect to $\sigma(\cX,\cX^\ast)$. It follows from $\sigma(\cX,\cX^\ast)$-lower semicontinuity that
\begin{equation}
\label{eq: reduction 1}
\varphi(X) \leq \liminf_{n\to\infty}\varphi(\E[X\,|\,\cF_n(X)]).
\end{equation}
Since for every $n\in\N$ we have $X\succeq_{cx}\E[X\,|\,\cF_n(X)]$ by Jensen's Inequality, we infer from the Schur convexity of $\varphi$ established in Theorem~\ref{theo: equivalence law invariance and schur convexity} that we also have
\begin{equation}
\label{eq: reduction 2}
\varphi(X) \geq \limsup_{n\to\infty}\varphi(\E[X\,|\,\cF_n(X)]).
\end{equation}
The desired assertion follows immediately by combining \eqref{eq: reduction 1} and \eqref{eq: reduction 2}.
\end{proof}

\smallskip

The next corollary records the companion statement for sets to the preceding restriction result. It shows that convex closed sets that are law invariant are completely determined by their intersection with $L^\infty$.

\begin{corollary}
For every convex, $\sigma(\cX,\cX^\ast)$-closed, and law-invariant set $\cA\subset\cX$ we have
\[
\cA=\cl_{\sigma(\cX,\cX^\ast)}(\cA\cap L^\infty).
\]
\end{corollary}
\begin{proof}
Clearly, we only need to show that $\cA\subset\cl_{\sigma(\cX,\cX^\ast)}(\cA\cap L^\infty)$. To this effect, take $X\in\cA$. It follows from Lemma~\ref{lem: density bounded} that we find a sequence $(\cF_n(X))$ of finitely-generated $\sigma$-fields in $\cF$ such that $\E[X\,|\,\cF_n(X)]\to X$ with respect to $\sigma(\cX,\cX^\ast)$. To conclude the proof, it suffices to observe that for every $n\in\N$ we have $X\succeq_{cx}\E[X\,|\,\cF_n(X)]$ by Jensen's Inequality and, hence, $\E[X\,|\,\cF_n(X)]\in\cA\cap L^\infty$ by Corollary~\ref{cor: equivalence law invariance schur sets}.
\end{proof}

\smallskip

It is easy to show that the conjugate functional of a quasiconvex and lower semicontinuous functional that is law invariant is itself law invariant. This allows us to derive a dual version of the preceding restriction result, which can be used to restrict the dual space to the space of bounded random variables. This has the following remarkable consequence, which extends \cite[Proposition 1]{Svindland2010} beyond the bounded setting. We also refer to \cite[Theorem 2.6]{ChenGaoXanthos2018}, \cite[Theorem 18]{LiebrichSvindland2019}, and \cite[Theorem 3.1]{LeungTantrawan2020} for equivalent formulations of this result in the setting of a rearrangement-invariant space.

\begin{theorem}
\label{theo: bounded dual}
For a proper, quasiconvex, law-invariant functional $\varphi:\cX\to[-\infty,\infty]$ the following statements are equivalent:
\begin{enumerate}[(a)]
  \item $\varphi$ is $\sigma(\cX,\cX^\ast)$-lower semicontinuous.
  \item $\varphi$ is $\sigma(\cX,L^\infty)$-lower semicontinuous.
\end{enumerate}
\end{theorem}
\begin{proof}
It suffices to establish the equivalence in the convex case. Indeed, the quasiconvex case can be established by converting the equivalence into a statement for convex sets via indicator functionals, see Corollary~\ref{cor: bounded dual}. So, assume that $\varphi$ is convex. Clearly, we only need to show that {\em (a)} implies {\em (b)}. To this effect, assume that $\varphi$ is $\sigma(\cX,\cX^\ast)$-lower semicontinuous. Recall that $\varphi^\ast$ is proper, convex, and $\sigma(\cX^\ast,\cX)$-lower semicontinuous. Moreover, for every $Y\in\cX^\ast$ we have
\[
\varphi^\ast(Y) = \sup_{X\in\cX}\{\E[XY]-\varphi(X)\} = \sup_{X\in\cX}\sup_{X'\sim X}\{\E[X'Y]-\varphi(X)\} = \sup_{X\in\cX}\left\{\int_0^1 q_X(s)q_Y(s)ds-\varphi(X)\right\}
\]
by Lemma~\ref{lem: chong rice}, showing that $\varphi^\ast$ is also law invariant. Now, fix $Y\in\cX^\ast$. It follows from Lemma~\ref{lem: density bounded} and Theorem~\ref{theo: restriction} that, for a suitable sequence $(\cF_n(Y))$ of finitely-generated $\sigma$-fields in $\cF$, we have $\E[Y\,\vert\,\cF_n(Y)]\to Y$ in the topology $\sigma(\cX^\ast,\cX)$ and $\varphi^\ast(\E[Y\,\vert\,\cF_n(Y)])\to\varphi^\ast(Y)$. This implies
\[
\E[XY]-\varphi^\ast(Y) = \lim_{n\to\infty}\E[X\E[Y\,\vert\,\cF_n(Y)]]-\varphi^\ast(\E[Y\,\vert\,\cF_n(Y)]) \leq \sup_{Z\in L^\infty}\{\E[XZ]-\varphi^\ast(Z)\}
\]
for every $X\in\cX$. As a result of Proposition~\ref{lem: fenchel moreau}, we infer that
\[
\varphi(X) = \sup_{Z\in L^\infty}\{\E[XZ]-\varphi^\ast(Z)\}
\]
for every $X\in\cX$. This shows that $\varphi$ is $\sigma(\cX,L^\infty)$-lower semicontinuous and concludes the proof.
\end{proof}

\smallskip

The preceding statement can be reformulated for sets by applying the above result to indicator functionals.

\begin{corollary}
\label{cor: bounded dual}
For a convex and law-invariant set $\cA\subset\cX$ the following statements are equivalent:
\begin{enumerate}[(a)]
  \item $\cA$ is $\sigma(\cX,\cX^\ast)$ closed.
  \item $\cA$ is $\sigma(\cX,L^\infty)$ closed.
\end{enumerate}
\end{corollary}


\section{Applications}

In this section we show how to exploit the restriction results established above to transfer a variety of known results about law-invariant or Schur-convex functionals on $L^\infty$ to the corresponding results on $\cX$. We refer to Theorem~\ref{theo: equivalence law invariance and schur convexity} for the equivalence between law invariance and Schur convexity.


\subsection{Quantile representations}

We start by extending the well-known quantile representation of law-invariant functionals. In the convex case, this extends \cite[Theorem 4.2]{Dana2005}, which was established in $L^\infty$, and \cite[Proposition 4.3]{GrechukZabarankin2012}, which was proved in $L^p$ under the additional requirement of positive homogeneity. It is worth pointing out that, even in the $L^p$ setting, our quantile representation is sharper because we can restrict the dual space to $L^\infty$. It is also worth noting that the arguments in \cite{GrechukZabarankin2012} exploit special properties of $L^p$ spaces, e.g.\ the norm density of $L^\infty$, that some our admissible choices for the space $\cX$, e.g.\ Orlicz spaces, do not necessarily satisfy. In the quasiconvex case, the following representation extends and complements \cite[Theorem 5.1]{CerreiaMaccheroniMarinacciMontrucchio2011}, which was established in $L^\infty$ under the additional assumption of monotonicity.

\begin{proposition}
\label{prop: quantile representation}
Let $\varphi:\cX\to[-\infty,\infty]$ be proper, $\sigma(\cX,\cX^\ast)$-lower semicontinuous, and law invariant.
\begin{enumerate}[(i)]
  \item If $\varphi$ is convex, then it can be represented as
\begin{equation}\label{quantile dual rep}
\varphi(X) = \sup_{Y\in L^\infty}\left\{\int_0^1 q_X(s)q_Y(s)ds-\varphi^\ast(Y)\right\}, \ \ \ X\in\cX,
\end{equation}
\[
\varphi^\ast(Y) = \sup_{X\in\cX}\left\{\int_0^1 q_X(s)q_Y(s)ds-\varphi(X)\right\}, \ \ \ Y\in\cX^\ast.
\]
  \item If $\varphi$ is quasiconvex, then it can be represented as
\[
\varphi(X) =
\sup_{Y\in L^\infty}F_\varphi^\ast\left(Y,\int_0^1 q_X(s)q_Y(s)ds\right), \ \ \ X\in\cX,
\]
\[
F_\varphi^\ast(Y,\alpha) = \inf\left\{\varphi(X) \,; \ X\in\cX, \ \int_0^1 q_X(1-s)q_Y(s)ds\geq\alpha\right\}, \ \ \ Y\in\cX^\ast,\,\alpha\in\R.
\]
\end{enumerate}
If $\varphi$ is additionally increasing, then we replace $L^\infty$ by $L^\infty_+$ in the above representations.
\end{proposition}
\begin{proof}
{\em (i)} It follows from Theorem~\ref{theo: bounded dual} that $\varphi$ is $\sigma(\cX,L^\infty)$-lower semicontinuous. Hence, we can apply Proposition~\ref{lem: fenchel moreau} to get
\begin{equation}
\label{eq: quantile representation 1}
\varphi(X)
=
\sup_{X'\sim X}\varphi(X')
=
\sup_{X'\sim X}\sup_{Y\in L^\infty}\{\E[X'Y]-\varphi^\ast(Y)\}
=
\sup_{Y\in L^\infty}\sup_{X'\sim X}\{\E[X'Y]-\varphi^\ast(Y)\}
\end{equation}
for every $X\in\cX$, where we used the law invariance of $\varphi$. Then, Lemma~\ref{lem: chong rice} yields
\[
\varphi(X) = \sup_{Y\in L^\infty}\left\{\int_0^1 q_X(s)q_Y(s)ds-\varphi^\ast(Y)\right\}
\]
for every $X\in\cX$. Similarly, we have
\[
\varphi^\ast(Y) = \sup_{X\in\cX}\{\E[XY]-\varphi(X)\} = \sup_{X\in\cX}\sup_{X'\sim X}\{\E[X'Y]-\varphi(X)\} = \sup_{X\in\cX}\left\{\int_0^1 q_X(s)q_Y(s)ds-\varphi(X)\right\}
\]
for every $Y\in\cX^\ast$. If $\varphi$ is additionally increasing, then the supremum over $L^\infty$ in~\eqref{eq: quantile representation 1} can be restricted to $L^\infty_+$ by Proposition~\ref{lem: fenchel moreau}.

\smallskip

{\em (ii)} It follows from Theorem~\ref{theo: bounded dual} that $\varphi$ is $\sigma(\cX,L^\infty)$-lower semicontinuous. Then, the law invariance of $\varphi$ and Proposition~\ref{lem: fenchel moreau} (note that we can restrict the supremum below to $L^\infty_+$ if $\varphi$ is increasing) imply that
\begin{equation}
\label{eq: quantile representation 2}
\varphi(X)
=
\sup_{X'\sim X}\varphi(X')
=
\sup_{X'\sim X}\sup_{Y\in L^\infty}F^\ast_\varphi(Y,\E[X'Y])
=
\sup_{Y\in L^\infty}\sup_{X'\sim X}F^\ast_\varphi(Y,\E[X'Y])
\end{equation}
for every $X\in\cX$. As $F^\ast_\varphi(Y,\cdot)$ is nondecreasing for every $Y\in L^\infty$, it follows from Lemma~\ref{lem: chong rice} that
\[
\varphi(X) = \sup_{Y\in L^\infty}F_\varphi^\ast\left(Y,\int_0^1 q_X(s)q_Y(s)ds\right)
\]
for every $X\in\cX$. Similarly, we have
\begin{align*}
F^\ast_\varphi(Y,\alpha) &=
\inf\{\varphi(X) \,; \ X\in\cX, \ \E[XY]\geq\alpha\} \\
&=
\inf\{\varphi(X) \,; \ X\in\cX, \ X'\sim X, \ \E[X'Y]\geq\alpha\} \\
&=
\inf\{\varphi(X) \,; \ X\in\cX, \ \inf_{X'\sim X}\E[X'Y]\geq\alpha\} \\
&=
\inf\left\{\varphi(X) \,; \ X\in\cX, \ \int_0^1 q_X(1-s)q_Y(s)ds\geq\alpha\right\}
\end{align*}
for all $Y\in\cX^\ast$ and $\alpha\in\R$. If $\varphi$ is additionally increasing, then the supremum over $L^\infty$ in~\eqref{eq: quantile representation 2} can be restricted to $L^\infty_+$ by Proposition~\ref{lem: fenchel moreau}.
\end{proof}

\smallskip

It follows from the above quantile representations that law invariance can be characterized at the level of conjugate functions for a quasiconvex and lower semicontinuous functional. In view of the equivalence between law invariance and Schur convexity, the same characterization holds by replacing law invariance with Schur convexity.

\begin{corollary}
\label{cor: law invariance and conjugate}
Let $\varphi:\cX\to[-\infty,\infty]$ be proper and $\sigma(\cX,\cX^\ast)$-lower semicontinuous.
\begin{enumerate}[(i)]
  \item If $\varphi$ is convex, then the following statements are equivalent:
\begin{enumerate}[(a)]
  \item $\varphi$ is law invariant.
  \item $\varphi^\ast$ is law invariant.
\end{enumerate}
  \item If $\varphi$ is quasiconvex, then the following statements are equivalent:
\begin{enumerate}[(a)]
  \item $\varphi$ is law invariant.
  \item $F_\varphi^\ast(\cdot,\alpha)$ is law invariant for every $\alpha\in\R$.
\end{enumerate}
\end{enumerate}
\end{corollary}
\begin{proof}
{\em (i)} It follows from Proposition~\ref{prop: quantile representation} that {\em (a)} implies {\em (b)}. Since $\varphi^\ast$ is proper, convex, and $\sigma(\cX^\ast,\cX)$-lower semicontinuous and $\varphi$ coincides with the conjugate of $\varphi^\ast$ by Proposition~\ref{lem: fenchel moreau}, we immediately obtain that {\em (b)} implies {\em (a)} again by Proposition~\ref{prop: quantile representation}.

\smallskip

{\em (ii)} It follows from Proposition~\ref{prop: quantile representation} that {\em (a)} implies {\em (b)}. To establish the converse implication, assume that $F_\varphi^\ast(\cdot,\alpha)$ is law invariant for every $\alpha\in\R$. Then, for every $X\in\cX$ we have
\[
\varphi(X) =
\sup_{Y\in\cX^\ast}F^\ast_\varphi(Y,\E[XY]) =
\sup_{Y\in\cX^\ast}\sup_{Y'\sim Y}F^\ast_\varphi(Y,\E[XY'])
\]
by Proposition~\ref{lem: fenchel moreau}. As $F^\ast_\varphi(Y,\cdot)$ is nondecreasing for every $Y\in\cX^\ast$, we infer from Lemma~\ref{lem: chong rice} that
\[
\varphi(X) = \sup_{Y\in\cX^\ast}F^\ast_\varphi\left(Y,\int_0^1 q_X(s)q_Y(s)ds\right)
\]
for every $X\in\cX$. This shows that $\varphi$ is law invariant and establishes that {\em (b)} implies {\em (a)}.
\end{proof}


\subsection{Extension results}

The next proposition shows that every convex and lower semicontinuous functional that is law invariant on $\cX$ can be uniquely extended to the entire space $L^1$ without losing its properties. In particular, $L^1$ can be viewed as the canonical model space for this type of functionals. In the convex case, this extension result was obtained in \cite[Theorem 2.2]{FilipovicSvindland2012} in the Lebesgue setting, in \cite[Theorem 1.4]{GaoLeungMunariXanthos2018} in the Orlicz setting, and in \cite[Theorem 2.6]{ChenGaoXanthos2018} in the rearrangement-invariant space setting. In the quasiconvex case, we refer to \cite[Proposition 20]{LiebrichSvindland2019} for a version in the rearrangement-invariant space setting.

\begin{proposition}
\label{prop: extension}
Let $\varphi:\cX\to[-\infty,\infty]$ be proper, $\sigma(\cX,\cX^\ast)$-lower semicontinuous, and law invariant.
\begin{enumerate}[(i)]
  \item If $\varphi$ is convex, then it can be uniquely extended to a proper, convex, $\sigma(L^1,L^\infty)$-lower semicontinuous, law-invariant functional $\overline{\varphi}:L^1\to[-\infty,\infty]$. The extension is explicitly given by
\[
\overline{\varphi}(X) = \sup_{Y\in L^\infty}
\left\{\int_0^1q_X(s)q_Y(s)ds-\gamma(Y)\right\}, \ \ \ X\in L^1,
\]
where
\[
\gamma(Y) = \sup_{X\in L^\infty}\left\{\int_0^1 q_X(s)q_Y(s)ds-\varphi(X)\right\}, \ \ \ Y\in L^\infty.
\]
  \item If $\varphi$ is quasiconvex, then it can be uniquely extended to a proper, quasiconvex, $\sigma(L^1,L^\infty)$-lower semicontinuous, law-invariant functional $\overline{\varphi}:L^1\to[-\infty,\infty]$. The extension is explicitly given by
\[
\overline{\varphi}(X) = \sup_{Y\in L^\infty}\Gamma\left(Y,\int_0^1q_X(s)q_Y(s)ds\right), \ \ \ X\in L^1,
\]
where
\[
\Gamma(Y,\alpha) = \inf\left\{\varphi(X) \,; \ X\in L^\infty, \ \int_0^1 q_X(1-s)q_Y(s)ds\geq\alpha\right\}, \ \ \ Y\in L^\infty,\,\alpha\in\R.
\]
\end{enumerate}
If $\varphi$ is additionally increasing, then $\overline{\varphi}$ is also increasing.
\end{proposition}
\begin{proof}
{\em (i)} Note that $\varphi_{|L^\infty}$ is proper, convex, $\sigma(L^\infty,\cX^\ast)$-lower semicontinuous, and law-invariant. In particular, $\varphi_{|L^\infty}$ is $\sigma(L^\infty,L^\infty)$-lower semicontinuous by Theorem~\ref{theo: bounded dual}. Then, Proposition~\ref{prop: quantile representation} yields
\begin{equation}
\label{eq: extension 1}
\varphi_{|L^\infty}(X) = \sup_{Y\in L^\infty}\left\{\int_0^1q_X(s)q_Y(s)ds-\gamma(Y)\right\}
\end{equation}
for every $X\in L^\infty$, where
\[
\gamma(Y) = \sup_{X\in L^\infty}\left\{\int_0^1 q_X(s)q_Y(s)ds-\varphi(X)\right\}
\]
for every $Y\in L^\infty$. Now, define the functional $\overline{\varphi}:L^1\to[-\infty,\infty]$ by setting
\[
\overline{\varphi}(X) = \sup_{Y\in L^\infty}\left\{\int_0^1 q_X(s)q_Y(s)ds-\gamma(Y)\right\} = \sup_{Y\in L^\infty}\sup_{Y'\sim Y}\{\E[XY']-\gamma(Y)\},
\]
where the last equality holds by Lemma~\ref{lem: chong rice}. It is clear that $\overline{\varphi}$ is a proper, convex, $\sigma(L^1,L^\infty)$-lower semicontinuous, and law-invariant extension of $\varphi_{|L^\infty}$. Note that $\overline{\varphi}_{|\cX}$ and $\varphi$ are both proper, convex, $\sigma(\cX,\cX^\ast)$-lower semicontinuous, and law-invariant extensions of $\varphi_{|L^\infty}$. Then, we must have $\varphi=\overline{\varphi}_{|\cX}$ by Theorem~\ref{theo: restriction}, so that $\overline{\varphi}$ is one of the desired extensions of $\varphi$. The uniqueness of such extensions follows immediately from Theorem~\ref{theo: restriction}. If $\varphi$ is additionally increasing, then the supremum in~\eqref{eq: extension 1} can be restricted to $L^\infty_+$ by Proposition~\ref{lem: fenchel moreau} and $\overline{\varphi}$ is easily seen to be increasing.

\smallskip

{\em (ii)} Note that $\varphi_{|L^\infty}$ is proper, quasiconvex, $\sigma(L^\infty,\cX^\ast)$-lower semicontinuous, and law-invariant. In particular, $\varphi_{|L^\infty}$ is $\sigma(L^\infty,L^\infty)$-lower semicontinuous by Theorem~\ref{theo: bounded dual}. Then, Proposition~\ref{prop: quantile representation} yields
\begin{equation}
\label{eq: extension 2}
\varphi_{|L^\infty}(X) = \sup_{Y\in L^\infty}\Gamma\left(Y,\int_0^1q_X(s)q_Y(s)ds\right)
\end{equation}
for every $X\in L^\infty$, where
\[
\Gamma(Y,\alpha) = \inf\left\{\varphi(X) \,; \ X\in L^\infty, \ \int_0^1 q_X(1-s)q_Y(s)ds\geq\alpha\right\}
\]
for all $Y\in L^\infty$ and $\alpha\in\R$. Now, define the functional $\overline{\varphi}:L^1\to(-\infty,\infty]$ by setting
\[
\overline{\varphi}(X) = \sup_{Y\in L^\infty}\Gamma\left(Y,\int_0^1q_X(s)q_Y(s)ds\right) = \sup_{Y\in L^\infty}\sup_{Y'\sim Y}\Gamma(Y',\E[XY']),
\]
where the last equality holds by Lemma~\ref{lem: chong rice}. It is clear that $\overline{\varphi}$ is a proper and law-invariant extension of $\varphi_{|L^\infty}$. Moreover, it follows from \cite[Lemma 2.7]{FrenkDiasGromicho1994} that $\overline{\varphi}$ is also quasiconvex and $\sigma(L^1,L^\infty)$-lower semicontinuous. Note that $\overline{\varphi}_{|\cX}$ and $\varphi$ are both proper, quasiconvex, $\sigma(\cX,\cX^\ast)$-lower semicontinuous, and law-invariant extensions of $\varphi_{|L^\infty}$. Then, we must have $\varphi=\overline{\varphi}_{|\cX}$ by Theorem~\ref{theo: restriction}, so that $\overline{\varphi}$ is one of the desired extensions of $\varphi$. The uniqueness of such extensions follows immediately from Theorem~\ref{theo: restriction}. If $\varphi$ is additionally increasing, then the supremum in~\eqref{eq: extension 2} can be restricted to $L^\infty_+$ by Proposition~\ref{lem: fenchel moreau} and $\overline{\varphi}$ is easily seen to be increasing.
\end{proof}

\smallskip

The next corollary features the counterpart for sets to the preceding extension results by showing that convex, $\sigma(\cX,\cX^\ast)$-closed, and law-invariant subsets of $\cX$ can always be retrieved from their norm closures in $L^1$. Here, we denote by $\cl_{\|\cdot\|_1}$ the closure with respect to the $L^1$ norm.

\begin{corollary}
For every convex, $\sigma(\cX,\cX^\ast)$-closed, and law-invariant set $\cA\subset\cX$ we have
\[
\cA=\cl_{\|\cdot\|_1}(\cA)\cap\cX.
\]
\end{corollary}
\begin{proof}
We only need to show that $\cl_{\|\cdot\|_1}(\cA)\cap\cX\subset\cA$. To this end, take $X\in\cX$ and assume that $X_n\to X$ with respect to the norm topology in $L^1$ for a suitable sequence $(X_n)\subset\cA$. In particular, $X_n\to X$ with respect to $\sigma(\cX,L^\infty)$. Since $\cA$ is $\sigma(\cX,L^\infty)$-closed by Corollary~\ref{cor: bounded dual}, we conclude that $X\in\cA$.
\end{proof}


\subsection{Dilatation monotonicity}

In this short section we recall the notion of dilatation monotonicity and show that it is equivalent to law invariance under quasiconvexity and lower semicontinuity. In the convex case, this extends \cite[Corollary 1.3]{ChernyGrigoriev2007} beyond the bounded setting and \cite[Theorem 2.1]{Svindland2014} beyond the Lebesgue setting. In the quasiconvex case, it extends \cite[Theorem 18]{LiebrichSvindland2019} beyond the setting of rearrangement-invariant spaces. We refer to \cite{RahseparXanthos2020} for a variety of results on dilatation monotonicity in the setting of general spaces.

\begin{definition}
A functional $\varphi:\cX\to[-\infty,\infty]$ is called {\em dilatation monotone} whenever
\[
\varphi(X) \geq \varphi(\E[X\,|\,\cG])
\]
for every $X\in\cX$ and every $\sigma$-field $\cG\subset\cF$ such that $\E[X\,|\,\cG]\in\cX$.
\end{definition}

\smallskip

\begin{proposition}
For a proper, quasiconvex, $\sigma(\cX,\cX^\ast)$-lower semicontinuous functional $\varphi:\cX\to[-\infty,\infty]$ the following statements are equivalent:
\begin{enumerate}[(a)]
  \item $\varphi$ is law invariant.
  \item $\varphi$ is dilatation monotone.
\end{enumerate}
\end{proposition}
\begin{proof}
Recall that $X\succeq_{cx}\E[X\,|\,\cG]$ for every $X\in\cX$ and every $\sigma$-field $\cG\subset\cF$ by Jensen's Inequality. Hence, it follows from Theorem~\ref{theo: equivalence law invariance and schur convexity} that {\em (a)} implies {\em (b)}. To show the converse implication, assume that $\varphi$ is dilatation monotone. We can easily follow the proof of Theorem~\ref{theo: restriction} to show that
\begin{equation}
\label{eq: approximation under dil mon}
\varphi(X) = \lim_{n\to\infty}\varphi_{|L^\infty}(\E[X\,|\,\cF_n(X)])
\end{equation}
for every $X\in\cX$ and for every increasing sequence $(\cF_n(X))$ of finitely-generated $\sigma$-fields in $\cF$ such that $\sigma(X)=\sigma(\bigcup_{n\in\N}\cF_n(X))$. Now, as $\varphi_{|L^\infty}$ is lower semicontinuous with respect to the norm topology on $L^\infty$, it follows from \cite[Theorem 2.1, Remark 2.2]{Svindland2014} that $\varphi_{|L^\infty}$ is law invariant. Now, take $X,Y\in\cX$ satisfying $X\sim Y$. Consider a refining sequence $(P_n)$ of finite partitions of $\R$ and set $\cF_n(X)=\sigma(\{X^{-1}(I) \,; \ I\in P_n\})$ and $\cF_n(Y)=\sigma(\{Y^{-1}(I) \,; \ I\in P_n\})$. Since $\E[X\,|\,\cF_n(X)]\sim\E[Y\,|\,\cF_n(Y)]$ for every $n\in\N$, we infer from~\eqref{eq: approximation under dil mon} that
\[
\varphi(X) = \lim_{n\to\infty}\varphi_{|L^\infty}(\E[X\,|\,\cF_n(X)]) = \lim_{n\to\infty}\varphi_{|L^\infty}(\E[Y\,|\,\cF_n(X)]) = \varphi(Y).
\]
This shows that $\varphi$ is law invariant and establishes that {\em (b)} implies {\em (a)}.
\end{proof}


\subsection{Infimal convolutions}

In this section we focus on infimal convolutions of law-invariant functionals. In the spirit of \cite{LiebrichSvindland2019}, we consider infimal convolutions with respect to general aggregation functions. In applications to finance and insurance, infimal convolutions appear naturally in the study of risk sharing and capital allocation problems, where the aggregation function is typically taken to be the sum or the maximum. We refer to the introduction for a list of references. In particular, we refer to \cite{ChenGaoXanthos2018} for a variety of results on classical infimal convolutions in the setting of a rearrangement-invariant space.

\begin{definition}
Let $n\in\N$ and $\Lambda:\R^n\to\R$ and for every $X\in\cX$ define the collection
\[
\cS_{\cX,n}(X) = \left\{(X_1,\dots,X_n)\in\cX^n \,; \ \sum_{i=1}^nX_i=X\right\}.
\]
Let $\varphi_1,\dots,\varphi_n:\cX\to[-\infty,\infty]$ be proper. The map $\square_{i=1}^n\varphi_i:\cX\to[-\infty,\infty]$ defined by
\[
\square_{i=1}^n\varphi_i(X) := \inf\{\Lambda(\varphi_1(X_1),\dots,\varphi_n(X_n)) \,; \ (X_1,\dots,X_n)\in\cS_{\cX,n}(X)\}
\]
is called the {\em ($\Lambda$-based) infimal convolution} of $\varphi_1,\dots,\varphi_n$.
\end{definition}

\smallskip

In line with the focus of our paper, we are interested in studying under which conditions the infimal convolution of quasiconvex, lower semicontinuous, and law-invariant functionals is still law invariant. This problem has been recently addressed in \cite{LiuWangWei2020} for standard sum-based infimal convolutions but without any quasiconvexity and lower semicontinuity assumptions. We show that, in our setting, law invariance is always preserved provided that the aggregation function is increasing and the space $\cX$ satisfies a suitable regularity property. We start with the following simple lemma about nonexpansive functions. Recall that a function $f:\R\to\R$ is said to be {\em nonexpansive} if $|f(x)-f(y)|\leq|x-y|$ for all $x,y\in\R$.

\begin{lemma}
\label{lem: nonexpansive}
Let $n\in\N$ and assume that $f_1,\dots,f_n:\R\to\R$ are increasing functions summing up to the identity function on $\R$. Then, $f_1,\dots,f_n$ are nonexpansive.
\end{lemma}
\begin{proof}
The statement is obvious if $n=1$, so assume that $n>1$. Set $g=\sum_{i=2}^nf_i$ and note that $g$ is increasing and $f_1$ and $g$ sum up to the identity function on $\R$. If $f_1$ is not nonexpansive, we find $x,y\in\R$ such that $x>y$ and $f_1(x)-f_1(y)>x-y$. In this case, however, we would get $g(y)-g(x)+x-y>x-y$, which violates the monotonicity of $g$. As a result, it follows that $f_1$ must be nonexpansive. Clearly, the same argument can be repeated for each of the other $f_i$'s.
\end{proof}

\smallskip

\begin{proposition}
\label{prop: law invariance inf conv}
Assume that $\Lambda$ is increasing and $f(X)\in\cX$ for every nonexpansive function $f:\R\to\R$ and every $X\in\cX$. Let $n\in\N$ and let $\varphi_1,\dots,\varphi_n:\cX\to[-\infty,\infty]$ be proper, quasiconvex, $\sigma(\cX,\cX^\ast)$-lower semicontinuous, and law invariant. Then, $\square_{i=1}^n\varphi_i$ is law invariant.
\end{proposition}
\begin{proof}
Take arbitrary $X,Y\in\cX$ such that $X\sim Y$. Clearly, it is enough to show that
\begin{equation}
\label{eq: inf conv}
\square_{i=1}^n\varphi_i(X) \geq \square_{i=1}^n\varphi_i(Y).
\end{equation}
To this effect, let $(X_1,\dots,X_n)\in\cS_{\cX,n}(X)$. The existence of comonotonic improvements established in \cite[Theorem 2]{LudkovskiRueschendorf2008} ensures that we find a comonotonic vector $(X_1^c,\dots,X_n^c)\in\cS_{L^1,n}(X)$ such that $X_i^c\preceq_{cx}X_i$ for every $i\in\{1,\dots,n\}$. It follows from the characterization of comonotonicity in \cite[Proposition 4.5]{Denneberg1994} that $X_1^c=f_1(X),\dots,X_n^c=f_n(X)$ for suitable increasing functions $f_1,\dots,f_n:\R\to\R$ summing up to the identity function on $\R$. In view of Lemma~\ref{lem: nonexpansive}, these functions are nonexpansive. In particular, by our assumption on $\cX$, we have that all the random variables $f_i(X)$'s and $f_i(Y)$'s belong to $\cX$. As a result,
\[
\Lambda(\varphi_1(X_1),\dots,\varphi_n(X_n))
\geq
\Lambda(\varphi_1(f_1(X)),\dots,\varphi_n(f_n(X)))
=
\Lambda(\varphi_1(f_1(Y)),\dots,\varphi_n(f_n(Y)))
\geq
\square_{i=1}^n\varphi_i(Y).
\]
Here, we used the monotonicity of $\Lambda$ and the Schur convexity established in Theorem~\ref{theo: equivalence law invariance and schur convexity} in the left-hand side inequality and law invariance in the equality. The desired inequality \eqref{eq: inf conv} now follows by taking the infimum over all $(X_1,\dots,X_n)\in\cS_{\cX,n}(X)$.
\end{proof}

\smallskip

\begin{remark}
If $\cX$ is an Orlicz space, then it is immediate to verify that $f(X)\in\cX$ for every nonexpansive function $f:\R\to\R$ and every $X\in\cX$. Indeed, it suffices to observe that $|f(X)|\leq|X|+|f(0)|$. More generally, this is true whenever $\cX$ contains all constant random variables and is solid in the sense that for all $X\in\cX$ and $Y\in L^0$ such that $|Y|\leq|X|$ it holds that $Y\in\cX$.
\end{remark}

\subsection{Dual representations of special risk measures}

In this section we extend a variety of dual representations obtained in the risk measure literature. We say that a functional $\varphi:\cX\to[-\infty,\infty]$ is {\em cash additive} if for all $X\in\cX$ and $m\in\R$ we have
\[
\varphi(X+m) = \varphi(X)-m.
\]
Moreover, we say that $\varphi$ is {\em comonotonic} if for all comonotone $X,Y\in\cX$ we have
\[
\varphi(X+Y)=\varphi(X)+\varphi(Y).
\]

\smallskip

Our representations will follow from the general extension principle recorded in the next lemma.

\begin{lemma}
\label{lem: lift up dual representations}
Let $\varphi:\cX\to[-\infty,\infty]$ be proper, convex, $\sigma(\cX,\cX^\ast)$-lower semicontinuous, and law invariant. Moreover, let $\cS$ be a set and assume there exist maps $\Phi:\cX\times\cS\to[-\infty,\infty]$ and $\gamma:\cS\to[-\infty,\infty]$ such that
\[
\varphi_{\vert L^\infty}(X) = \sup_{S\in\cS}\{\Phi(X,S)-\gamma(S)\}, \ \ \ X\in L^\infty.
\]
If $\Phi(\cdot,S)$ is convex, $\sigma(\cX,\cX^\ast)$-lower semicontinuous, and law invariant for every $S\in\cS$, then
\[
\varphi(X) = \sup_{S\in\cS}\{\Phi(X,S)-\gamma(S)\}, \ \ \ X\in\cX.
\]
\end{lemma}
\begin{proof}
We claim that $\varphi$ coincides with the functional $\psi:\cX\to[-\infty,\infty]$ defined by
\[
\psi(X) = \sup_{S\in\cS}\{\Phi(X,S)-\gamma(S)\}, \ \ \ X\in\cX.
\]
To show this, note that $\psi$ is convex, $\sigma(\cX,\cX^\ast)$-lower semicontinuous, and law invariant by our assumptions on $\Phi$. Note also that $\varphi_{\vert L^\infty}$ is proper by properness of $\varphi$ and by Theorem~\ref{theo: restriction}. If $\psi(X)=-\infty$ for some $X\in\cX$, then \cite[Proposition 2.2.5]{Zalinescu2002} would imply that $\psi$ could not take any finite value, which is impossible because $\psi_{\vert L^\infty}=\varphi_{\vert L^\infty}$. Hence, we must have $\psi(X)>-\infty$ for every $X\in\cX$. The properness of $\varphi_{\vert L^\infty}$ now implies that $\psi$ is also proper. As $\varphi$ and $\psi$ coincide on $L^\infty$, we must have $\varphi=\psi$ by Theorem~\ref{theo: restriction}, concluding the proof.
\end{proof}

\smallskip

As a first application, we use the above extension principle to derive a general formulation of the Kusuoka representation. This representation was first established in the space of bounded random variables by \cite{Kusuoka2001}, under the assumption of positive homogeneity, and by \cite{FrittelliRosazza2005}. Later, it was extended to a general Lebesgue setting by \cite{Shapiro2013}, again under the assumption of positive homogeneity, and to a general Orlicz setting by \cite{GaoLeungMunariXanthos2018}. Here, for every $s\in[0,1]$ and every $X\in L^1$ we define the {\em Expected Shortfall} of $X$ at level $s$ by
\[
\ES_s(X) :=
\begin{cases}
-\frac{1}{s}\int_0^s q_X(t)dt & \mbox{if} \ s\in(0,1],\\
-\essinf(X) & \mbox{if} \ s=0.
\end{cases}
\]
In particular, note that $\ES_1(X)=\E[-X]$. Moreover, we denote by $\cP((0,1])$ the set of probability measures on $(0,1]$ (equipped with the Borel $\sigma$-field).

\begin{proposition}
\label{prop: kusuoka}
Assume that $\cX$ is a rearrangement-invariant space. Let $\varphi:\cX\to[-\infty,\infty]$ be proper, convex, $\sigma(\cX,\cX^\ast)$-lower semicontinuous, law invariant, decreasing, and cash additive. Then, we have
\[
\varphi(X) = \sup_{\mu\in\cP((0,1])}\left\{\int_{(0,1]}\ES_s(X)d\mu(s)-\gamma(\mu)\right\}, \ \ \ X\in\cX,
\]
where
\[
\gamma(\mu) = \sup\left\{\int_{(0,1]}\ES_s(X)d\mu(s) \,; \ X\in L^\infty, \ \varphi(X)\leq0\right\}, \ \ \ \mu\in\cP((0,1]).
\]
\end{proposition}
\begin{proof}
Note that for all $X\in\cX$ and $\mu\in\cP((0,1])$ the integral $\int_{(0,1]}\ES_s(X)d\mu(s)$ is well defined because $\ES_s(X)\geq\E[-X]$ for every $s\in(0,1]$. It follows from \cite[Theorem 4.62]{FoellmerSchied2011} that
\[
\varphi_{|L^\infty}(X) = \sup_{\mu\in\cP((0,1])}\left\{\int_{(0,1]}\ES_s(X)d\mu(s)-\gamma(\mu)\right\}
\]
for every $X\in L^\infty$, where
\[
\gamma(\mu) = \sup\left\{\int_{(0,1]}\ES_s(X)d\mu(s) \,; \ X\in L^\infty, \ \varphi(X)\leq0\right\}
\]
for every $\mu\in\cP((0,1])$. Now, consider the functional $\Phi:\cX\times\cP((0,1])\to(-\infty,\infty]$ given by
\[
\Phi(X,\mu) = \int_{(0,1]}\ES_s(X)d\mu(s).
\]
It is clear that, for every $\mu\in\cP((0,1])$, the functional $\Phi(\cdot,\mu)$ is convex and law invariant. We claim that $\Phi(\cdot,\mu)$ satisfies the Fatou property. To see this, take a sequence $(X_n)\subset\cX$ and $X\in\cX$ such that $X_n\to X$ almost surely and $\sup_{n\in\N}|X_n|\in\cX$. Set $Y=\sup_{n\in\N}|X_n|$. Since $\ES_s$ has the Fatou property for every $s\in(0,1]$ and $\ES_s(X_n)\geq\ES_s(Y)\geq\E[-Y]$ for all $s\in(0,1]$ and $n\in\N$, we get
\[
\Phi(X,\mu) = \int_{(0,1]}\lim_{n\to\infty}\ES_s(X_n)d\mu(s) \leq \liminf_{n\to\infty}\int_{(0,1]}\ES_s(X_n)d\mu(s) = \liminf_{n\to\infty}\Phi(X_n,\mu)
\]
by Fatou's Lemma. This establishes the Fatou property. In view of this, Proposition~\ref{prop: fatou} implies that $\Phi(\cdot,\mu)$ is $\sigma(\cX,\cX^\ast)$-lower semicontinuous. The desired representation now follows from Lemma~\ref{lem: lift up dual representations}.
\end{proof}

\smallskip

Next, we present a general version of the Kusuoka representation in the case of comonotonic risk measures. We denote by $\cP([0,1])$ the set of probability measures on $[0,1]$ (equipped with the Borel $\sigma$-field).

\begin{proposition}
Assume that $\cX$ is a rearrangement-invariant space. Let $\varphi:\cX\to[-\infty,\infty]$ be proper, convex, $\sigma(\cX,\cX^\ast)$-lower semicontinuous, law invariant, decreasing, cash additive, and comonotonic. Then, there exists $\mu\in\cP([0,1])$ such that
\[
\varphi(X) =\int_{[0,1]}\ES_s(X)d\mu(s), \ \ \ X\in\cX.
\]
\end{proposition}
\begin{proof}
Note that for all $X\in\cX$ and $\mu\in\cP([0,1])$ the integral $\int_{[0,1]}\ES_s(X)d\mu(s)$ is well defined because $\ES_s(X)\geq\E[-X]$ for every $s\in[0,1]$. It follows from \cite[Theorem 4.93]{FoellmerSchied2011} that there exists $\mu\in\cP([0,1])$ such that
\[
\varphi_{|L^\infty}(X) = \int_{[0,1]}\ES_s(X)d\mu(s)
\]
for every $X\in L^\infty$. Now, consider the functional $\Phi:\cX\to(-\infty,\infty]$ given by
\[
\Phi(X) = \int_{[0,1]}\ES_s(X)d\mu(s).
\]
We can argue as in the proof of Proposition~\ref{prop: kusuoka} to show that $\Phi$ is convex, $\sigma(\cX,\cX^\ast)$-lower semicontinuous, and law invariant. As a result, the desired representation follows immediately from Lemma~\ref{lem: lift up dual representations}.
\end{proof}

\smallskip

We conclude by providing a dual representation of risk measures based on loss functions that extends \cite[Theorem 4.115]{FoellmerSchied2011} beyond the bounded setting. A function $\ell:\R\to\R$ is called a {\em loss function} if it is nonconstant, nondecreasing, and convex. We denote by $\cP^\infty$ the set of probability measures on $(\Omega,\cF)$ that are absolutely continuous with respect to $\probp$ and satisfy $\frac{d\probq}{d\probp}\in L^\infty$.

\begin{proposition}
Assume that $\cX$ is a rearrangement-invariant space. Let $\ell$ be a loss function and take $\ell_0\in\R$ such that $\ell(x_0)<\ell_0$ for some $x_0\in\R$. The functional $\varphi_\ell:\cX\to[-\infty,\infty]$ defined by
\[
\varphi_\ell(X) = \inf\{m\in\R \,; \ \E[\ell(-X-m)]\leq\ell_0\}, \ \ \ X\in\cX,
\]
satisfies
\[
\varphi_\ell(X) = \sup_{\probq\in\cP^\infty}\left\{\E_\probq[-X]-\inf_{\lambda\in(0,\infty)}
\frac{1}{\lambda}\left(\ell_0+\E\left[\ell^\ast\left(\lambda\frac{d\probq}{d\probp}\right)\right]\right)\right\}, \ \ \ X\in\cX,
\]
where
\[
\ell^\ast(y) = \sup_{x\in\R}\{xy-\ell(x)\}, \ \ \ y\in\R.
\]
\end{proposition}
\begin{proof}
First of all, note that $\E[\ell(X)]$ is well defined for every $X\in\cX$ by convexity of $\ell$. Since $\E[\ell(X)]\geq\ell(\E[X])$ for every $X\in\cX$ by Jensen's inequality and $\ell_0$ belongs to the interior of the range of $\ell$ by assumption, we have $\varphi_\ell(X)>-\infty$ for every $X\in\cX$. Moreover, $\varphi_\ell(0)\leq-x_0<\infty$. This shows that $\varphi_\ell$ is proper. It is clear that $\varphi_\ell$ is convex and law invariant. We claim that $\varphi_\ell$ satisfies the Fatou property. To prove this, it suffices to show that for every sequence $(X_n)\subset\cX$ and every $X\in\cX$ such that $X_n\to X$ almost surely and $\sup_{n\in\N}|X_n|\in\cX$ we have
\begin{equation}
\label{eq: loss}
\sup_{n\in\N}\E[\ell(-X_n)]\leq\ell_0 \ \implies \ \E[\ell(-X)]\leq\ell_0.
\end{equation}
To establish \eqref{eq: loss}, note that $\ell(X_n)\to\ell(X)$ almost surely by continuity of $\ell$. Set $Y=\sup_{n\in\N}|X_n|$. By convexity of $\ell$ we find $a,b\in\R$ such that $\ell(0)\geq\ell(-Y)\geq a(-Y)+b$. This implies that $\ell(-Y)$ is integrable. Since $\ell(-X_n)\geq\ell(-Y)$ for every $n\in\N$, we can apply Fatou's Lemma to obtain
\[
\E[\ell(-X)] = \E\left[\lim_{n\to\infty}\ell(X_n)\right] \leq \liminf_{n\to\infty}\E[\ell(X_n)] \leq \ell_0.
\]
This yields \eqref{eq: loss} and proves that $\varphi_\ell$ satisfies the Fatou property. As a result, Proposition~\ref{prop: fatou} implies that $\varphi_\ell$ is $\sigma(\cX,\cX^\ast)$-lower semicontinuous. Since $\varphi_\ell$ is also $\sigma(\cX,L^\infty)$-lower semicontinuous by Theorem~\ref{theo: bounded dual}, we can argue as in \cite[Theorem 4.115]{FoellmerSchied2011} (see also \cite[Theorem 5.1]{ArducaKochMunari2019} for a simpler proof) to get
\[
(\varphi_\ell)_{|L^\infty}(X) = \sup_{\probq\in\cP^\infty}\left\{\E_\probq[-X]-\inf_{\lambda\in(0,\infty)}
\frac{1}{\lambda}\left(\ell_0+\E\left[\ell^\ast\left(\lambda\frac{d\probq}{d\probp}\right)\right]\right)\right\}
\]
for every $X\in L^\infty$. The desired representation on $\cX$ is now a direct consequence of Theorem~\ref{theo: restriction}.
\end{proof}

\end{document}